\documentclass[notoc,12pt,twoside,nohyper]{JHEP3} 
\usepackage{epsfig}
\usepackage{amssymb}
\usepackage{amsmath}
\usepackage{amsfonts}
\usepackage{amssymb}
\usepackage{amscd}
\usepackage{amsthm}
\usepackage{graphicx}
\newcommand\fverb{\setbox\pippobox=\hbox\bgroup\verb}
\newcommand\fverbdo{\egroup\medskip\noindent%
            \fbox{\unhbox\pippobox}\ }
\newcommand\fverbit{\egroup\item[\fbox{\unhbox\pippobox}]}
\newbox\pippobox
\title{Scale dependence of heavy quark production in pp and
$\overline{\mathbf{p}}$p collisions - transverse momentum dependence.}

\author{by J. Srbek and J. Ch\'{y}la
\\
    Center of Particle Physics,
    Institute of Physics AS CR, Na Slovance 2, Prague 8, Czech Republic\\
    E-mail: \email{chyla@fzu.cz, srbekj@fzu.cz}}

\abstract{The dependence of the differential cross section
${\mathrm{d}\sigma}/{\mathrm{d}p_{\perp}}$ of inclusive heavy
quark production in pp and $\overline{\mathrm{p}}$p collisions on the
renormalization and factorization scales is investigated. The implications
of our results for experiments at TEVATRON and LHC are discussed. In
particular, it is shown that the NLO QCD predictions for $\overline{t}t$
production at the LHC based on the Principle of Minimal Sensitivity
are by $30-50$\% higher than the standard ones.}


\begin{document}

\maketitle 

\section{Introduction}
\label{intro}
Heavy quark production in hard collisions of hadrons, leptons and
photons has been considered as a clean test of perturbative
QCD. It has therefore come as a surprise that the first data on the
$\overline{b}b$ production in $\overline{\mathrm{p}}$p collisions at the
TEVATRON $\gamma$p collisions at HERA and $\gamma\gamma$ collisions at
LEP2 have turned out to lie significantly and systematically above
theoretical calculations. The disagreement between data and theory
is particularly puzzling for the collisions of two quasireal photons
at LEP2.

The NLO QCD calculations of heavy quark production in these collisions
depend on a number of inputs: $\alpha_s$, parton distribution functions
(PDF) of colliding hadrons or (resolved) photons, fragmentation functions,
masses of heavy quarks, and last but not least the choice of
renormalization (RS) and factorization (FS) scales $\mu$ and $M$.

The discrepancy between the D0 \cite{D0data}
and CDF data \cite{CDFdata} on $\overline{b}b$ production
and theoretical calculations \cite{mangano} has led to suggestions that
it might represent the manifestation of effects of unintegrated gluon
distribution function within the $k_T$-factorization approach
\cite{hannes}, or even a signal of supersymmetry \cite{susy2}. In both
cases the agreement between the TEVATRON data
and the corresponding calculations \cite{hannes,susy2} is quite
impressive. On the other hand, it has been argued in \cite{nason2} that
proper parameterization of the b-quark fragmentation functions also
improves the agreement between CDF data and
QCD calculations. There have furthermore been attempts to go beyond
fixed order calculations by resumming the effects of large
logs of the type $\ln(S/4m_b^2)$ \cite{catani}, where $S$ denotes the
center-of-mass energy squared, or $\ln(p_{\perp}/m_b)$
\cite{cacciari}, as well as the effects of soft gluon resummation
important in the threshold region \cite{bonciani}.

In \cite{jch} the sensitivity of NLO QCD calculations of the total
cross section $\sigma_{tot}(p\overline{p}\rightarrow b\overline{b})$
to the variation of the renormalization and factorization scales $\mu$
and $M$ was investigated in some detail. It was shown that in order to
arrive at locally stable results these two scales must be kept
independent. Moreover, it turned out that in the TEVATRON energy range
the position of the saddle point of the cross section
$\sigma_{tot}^{\mathrm{NLO}}(p\overline{p}\rightarrow b\overline{b};S,M,\mu)$,
considered as a function of $\mu$ and $M$, lies far away from the diagonal
$\mu=M$. Using the NLO prediction at
the saddle point instead of the conventional choice $\mu=M=m_b$ enhances the
theoretical prediction in the TEVATRON energy range by a factor of about 2,
which may help in explaining part of the excess of data over NLO QCD
predictions.

In this paper the analysis of \cite{jch} is extended to the
differential cross section ${\mathrm{d}\sigma}/{\mathrm{d}}p_{\perp}$ of
inclusive heavy quark production at the NLO.
Bottom as well as top quark production will
be considered in circumstances relevant for experiments at TEVATRON and LHC.

\vspace*{-0.3cm}
\section{Basic facts and formulae}
\label{basics}
The basic quantity of perturbative QCD calculations, the renormalized color
coupling $\alpha_s(\mu)$, depends on the renormalization scale $\mu$ in a
way governed by the equation
\begin{equation}
\frac{{\mathrm d}\alpha_s(\mu)}{{\mathrm d}\ln \mu^2}\equiv
\beta(\alpha_s(\mu))=
-\frac{\beta_0}{4\pi}\alpha_s^2(\mu)-
\frac{\beta_1}{16\pi^2}
\alpha_s^3(\mu)+\cdots,
\label{RG}
\end{equation}
where for $n_f$ massless quarks $\beta_0=11-2n_f/3$ and $\beta_1=102-38n_f/3$.
The solutions of (\ref{RG}) depend beside $\mu$ also on the renormalization
scheme (RS). At the NLO (i.e. taking into account first two terms in
(\ref{RG})) this RS can be specified, for instance, via the parameter
$\Lambda_{\mathrm{RS}}$, corresponding to the renormalization scale for
which $\alpha_s$ diverges. The coupling $\alpha_(\mu)$ is then given as the
solution of the equation
\begin{equation}
\frac{\beta_0}{4\pi}\ln\left(\frac{\mu^2}{\Lambda^2_{\mathrm{RS}}}\right)=
\frac{1}{\alpha_s(\mu)}+
c\ln\frac{c\alpha_s(\mu)}{1+c\alpha_s(\mu)},
\label{equation}
\end{equation}
where $c=\beta_1/(4\pi\beta_0)$. At the NLO the coupling $\alpha_s$ is thus
a function of the ratio $\mu/\Lambda_{\mathrm{RS}}$ and the variation of the
RS for fixed scale $\mu$ is therefore equivalent to the variation of $\mu$
for fixed RS. To vary both the renormalization scale and scheme is legitimate,
but redundant. Let us emphasize that the choice of the RS is as important as
the choice of renormalization scale. As setting the renormalization scale equal
to some ``natural physical scale'' gives different results in different RS, the
existence of such a ``natural scale'' is actually of no help in arriving at
unique result. If not stated otherwise, we shall work
in the conventional $\overline{\mathrm{MS}}$ RS and vary the renormalization
scale $\mu$ only.

For hadrons the factorization scale dependence of PDF is determined by the
system of evolution equations for quark singlet, nonsinglet and gluon
distribution functions
\begin{eqnarray}
\frac{{\mathrm d}\Sigma(M)}{{\mathrm d}\ln M^2}& =&
P_{qq}(M)\otimes \Sigma(M)+ P_{qG}(M)\otimes G(M),
\label{Sigmaevolution}
\\ \frac{{\mathrm d}G(M)}{{\mathrm d}\ln M^2} & =&
P_{Gq}(M)\otimes \Sigma(M)+ P_{GG}(M)\otimes G(M), \label{Gevolution} \\
\frac{{\mathrm d}q_{\mathrm {NS}}(M)}{{\mathrm d}\ln M^2}& =&
P_{\mathrm {NS}}(M)\otimes q_{\mathrm{NS}}(M),
\label{NSevolution}
\end{eqnarray}
where
\begin{eqnarray}
\Sigma(x,M) & \equiv &
\sum_{i=1}^{n_f} \left(q_i(x,M)+\overline{q}_i(x,M)\right),
\label{singlet}\\
q_{\mathrm{NS},i}(x,M)& \equiv &(q_i(x,M)+\overline{q}_i(x,M))-
\frac{1}{n_f}\Sigma(x,M),~\forall i.
\label{nonsinglet}
\end{eqnarray}
The splitting functions admit expansion in powers of $\alpha_s(M)$
\begin{equation}
P_{ij}(x,M)=\frac{\alpha_s(M)}{2\pi}P^{(0)}_{ij}(x) +
\left(\frac{\alpha_s(M)}{2\pi}\right)^2 P_{ij}^{(1)}(x)+\cdots,
\label{splitpij}
\end{equation}
where $P^{(0)}_{ij}(x)$ are {\em unique}, whereas all higher order
splitting functions $P^{(j)}_{kl},j\ge 1$ depend on the choice of the
factorization scheme (FS). The equations
(\ref{Sigmaevolution}-\ref{NSevolution}) can be recast into evolution
equations for $q_i(x,M),\overline{q}_i(x,M)$ and $G(x,M)$.
In all calculations presented below we took $n_f=4$ in the case of bottom quark
and $n_f=5$ for top quark. The RG equation (\ref{equation}) was solved
numerically in order to guarantee correct behaviour of its solution for small
renormalization scales.

\section{General form of the cross section}
According to the factorization theorem, the differential cross section for
inclusive $Q\overline{Q}$ production in hadronic collisions has the form
(suppressing the specification of the dependence on $m_Q$ and $\sqrt{S}$)
\begin{equation}
\frac{\mathrm{d}\sigma(p_\perp,y)}{\mathrm{d}y\mathrm{d}p_\perp}=
\sum_{i,j=q,\bar{q},G}
\int\!\!\!\int\mathrm{d}x_1\mathrm{d}x_2f_{i/A}(x_1,M)f_{j/B}(x_2,M)
\frac{\mathrm{d}
\sigma_{ij}(x_1x_2,M, p_\perp,y)}
{\mathrm{d}y\mathrm{d}p_\perp},
\label{xs}
\end{equation}
where $y$ and $p_{\perp}$ denote the rapidity and transverse momentum of a
heavy quark $Q$, $A$ and $B$ stand for the beam particles and the partonic
hard scattering cross sections admit expansions in powers of $\alpha_s(\mu)$
\begin{equation}
\frac{{\mathrm{d}}\sigma_{ij}(x_1x_2,M,p_{\perp},y)}
{{\mathrm{d}}y{\mathrm{d}}p_{\perp}}=
\alpha_s^2(\mu)c_{ij}^{(2)}(x_1x_2,p_\perp,y)
+\alpha_s^3(\mu)c_{ij}^{(3)}(x_1x_2,M,\mu,p_\perp,y)+\cdots.
\label{expansion}
\end{equation}
Note that whereas the
factorization scale scale $M$ appears as argument of both the PDF $f_{i/A},f_{j/B}$
and parton level hard scattering cross sections $\sigma_{ij}$, the renormalization
scale $\mu$ enters only if we calculate the latter as perturbative expansion in
powers of $\alpha_s$. Both the renormalization and factorization scales are auxiliary,
unphysical quantities, on which observables like the cross section (\ref{xs}) cannot
depend, provided both the expansions (\ref{expansion}) and (\ref{splitpij}) are
taken to all orders. In such a case any choice of $\mu,M$ gives the same
result for (\ref{xs}). If, however, these expansions are considered to a finite
order, as is in practice always the case, the expression on the right hand
side of (\ref{xs}) develops nontrivial dependence on these two scales and their
choice therefore matters. Moreover, as the renormalization scale
reflects ambiguity in the treatment of short distances, whereas the factorization
scale comes from similar ambiguity concerning large distances, these two scales
should be kept as independent free parameters of any finite order
perturbative approximations \cite{politzer}.

\section{Choice of scales}
\label{choice}
For a given $p_{\perp}$, the result of the evaluation of (\ref{xs}) to the
NLO can be represented graphically in the form of two-dimensional surface over
the plane $\mu,M$. In this representation any of the points on such a surface
corresponds to one, in principle equally legitimate, choice of these two scales.
By constructing this surface we get quantitative idea of the stability of our
calculations. Specifically we can look for the presence of saddle points,
where the NLO results exhibit locally the independence of $\mu$ and $M$,
which is a necessary feature of all-order calculations globally. The choice
of scales based on this criterion, called Principle of Minimal Sensitivity
(PMS)~\cite{pms}, is one of the two general approaches to scale setting, the
other being the Effective Charges method \cite{EC}. This latter approach selects
those points $\mu_E,M_E$ for which the LO and NLO approximations of (\ref{xs})
coincide and the perturbative expansion thus has the ``fastest apparent
convergence''. Graphically this choice of scales is given by the intersection of
LO and NLO surfaces representing the cross section
${\mathrm{d}}\sigma/{\mathrm{d}}p_\perp(\mu,M)$.
As shown in \cite{jch} for the total cross section the resulting points form a
curve which passes very close to the saddle point, if such a point does exist.
As will be shown below the same holds for differential cross section
${\mathrm{d}}\sigma/{\mathrm{d}}p_\perp$.

In this connection we want to recall the simple but often overlooked fact that
the conventional procedure of setting $\mu=M$ and identifying it with some
``natural physical scale'' (in our case usually $m_{\perp}\equiv\sqrt{m_Q^2+
p_{\perp}^2}$) does not actually resolve the underlying ambiguity, because it
implicitly assumes that we are working in the $\overline{\mathrm{MS}}$ RS. There
is, however, nothing ``natural'' on this (or any other) RS, as the
$\overline{\mathrm{MS}}$ is just one of the infinite number of equally legitimate
choices of the RS. As emphasized in Section 2, working in a different RS of the
couplant $\alpha_s(\mu)$, the same choice of ``natural'' renormalization scale
$\mu$ leads to a different result for the cross section (\ref{xs}). This fact
also implies that it is meaningless to estimate the ``theoretical uncertainty''
related to freedom in the choice of the renormalization scale $\mu$ by varying
it in some interval around the ``natural'' scale because the same procedure
leads in different RS to different bands of theoretical predictions. The standard
choice of $\overline{\mathrm{MS}}$ RS is often justified by noting that it leads
to small coefficients in perturbative expansions of physical cross sections. But
if the apparent rate of convergence is the basic criterion for the ``proper''
choice of the scales than the points preferred by the EC approach, rather than
those defining the $\overline{\mathrm{MS}}$ RS should be selected.

On the other hand, although the positions of the saddle points do also depend
on the chosen RS, the values of the cross section (\ref{xs}) at these saddle
points do not! Contrary to the scale setting methods relying on the existence of
``natural physical scale'' the PMS approach therefore results in a unique
prediction. For predictions based on the method of effective charges \cite{EC}
the same uniqueness holds for any fixed factorization scale $M$ as well. In the
case both $\mu$ and $M$ are treated as independent free parameter the EC approach
does not yield a unique prediction but the whole continuum of values reflecting the
fact that the intersection of LO and NLO surfaces defines a curve
\footnote{This statement should be taken with some reservation as there
is in fact no guarantee that the saddle points and/or the mentioned
intersection do, in a given kinematic region, exist.}.

However, independently of the strategy one prefers for choosing the
renormalization and factorization scales, it is undoubtedly useful to have
quantitative idea of the sensitivity of our calculations to the
variation of these unphysical parameters.

\section{Results}
\label{results}
The cross section (\ref{xs}) was computed to the NLO using the code of the ref.
\cite{mangano} together with CTEQ6.1 set of PDF (with $\Lambda_4=326$~MeV). As no
restrictions on the rapidity $y$ of the produced heavy quark was imposed in this
study, the integration over the whole kinematically allowed region of $y$ was
performed. To see how much our conclusions depend on the knowledge of PDF, some
of the calculations were done also for the GRV94 set. The difference between the
CTEQ6.1 and GRV94 results, should, however, not be taken as an estimate of their
dependence of the chosen PDF. We have deliberately taken the now obsolete GRV94
set in order to illustrate the independence of the basic features of the
renormalization and factorization scale dependence
of $\mathrm{d}\sigma^{\mathrm{NLO}}/{\mathrm{d}}   p_\perp$ on the selected PDF.

The numerical integration used in \cite{mangano} employs standard Vegas method.
For the evaluation of the LO contribution 4 iterations each with 4000 points were
used, whereas for the NLO term these numbers were increased to 6 iterations each
with 8000 points.

\subsection{$\overline{b}b$ production}
\label{b}
For inclusive $\overline{b}b$ production the differential cross section
${\mathrm{d}}\sigma/{\mathrm{d}}p_\perp$ was computed for $\overline{\mathrm{p}}$p
collisions at $\sqrt{S}=64, 630$ and $1800$ GeV and for pp collisions at
$\sqrt{S}=14$ TeV. For $\overline{\mathrm{p}}$p collisions the
first energy does not correspond to any real experiment, but we have included it
to make contact with the earlier studies of this kind \cite{guido}. In all
calculations we set $m_b=4.95$ GeV.
\begin{figure}[b]
\includegraphics[width=4cm]{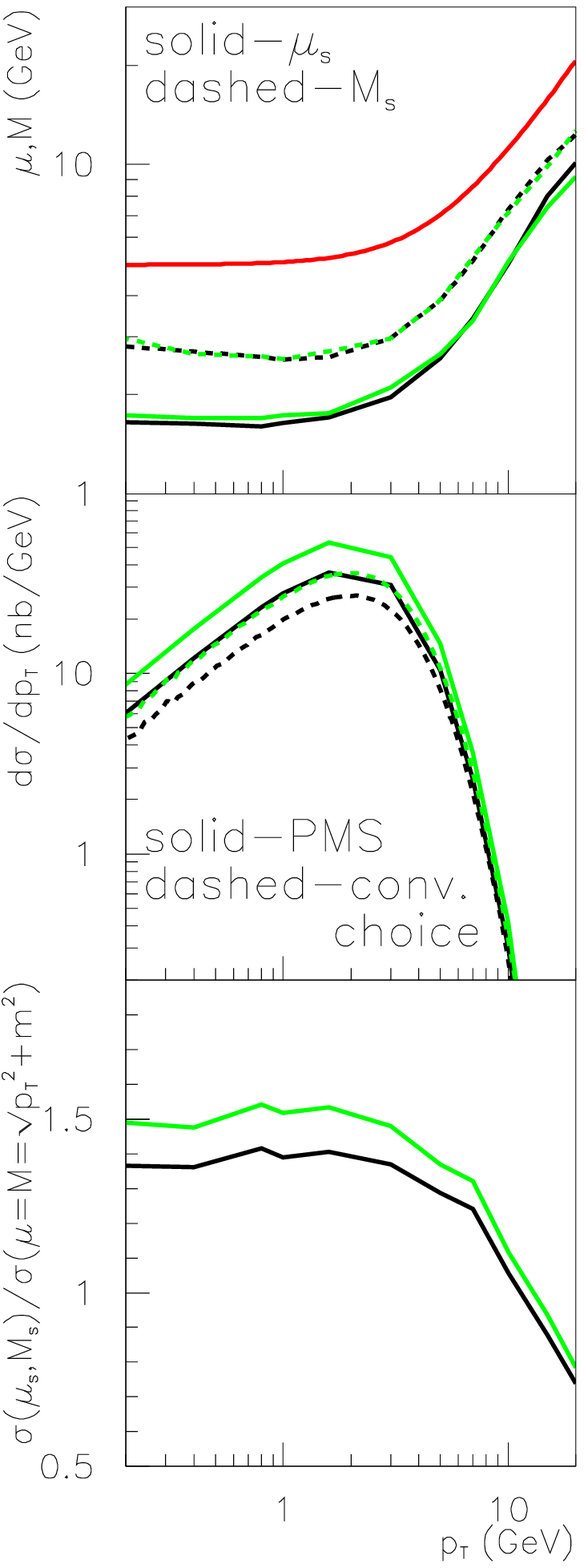}~
\includegraphics[width=11cm]{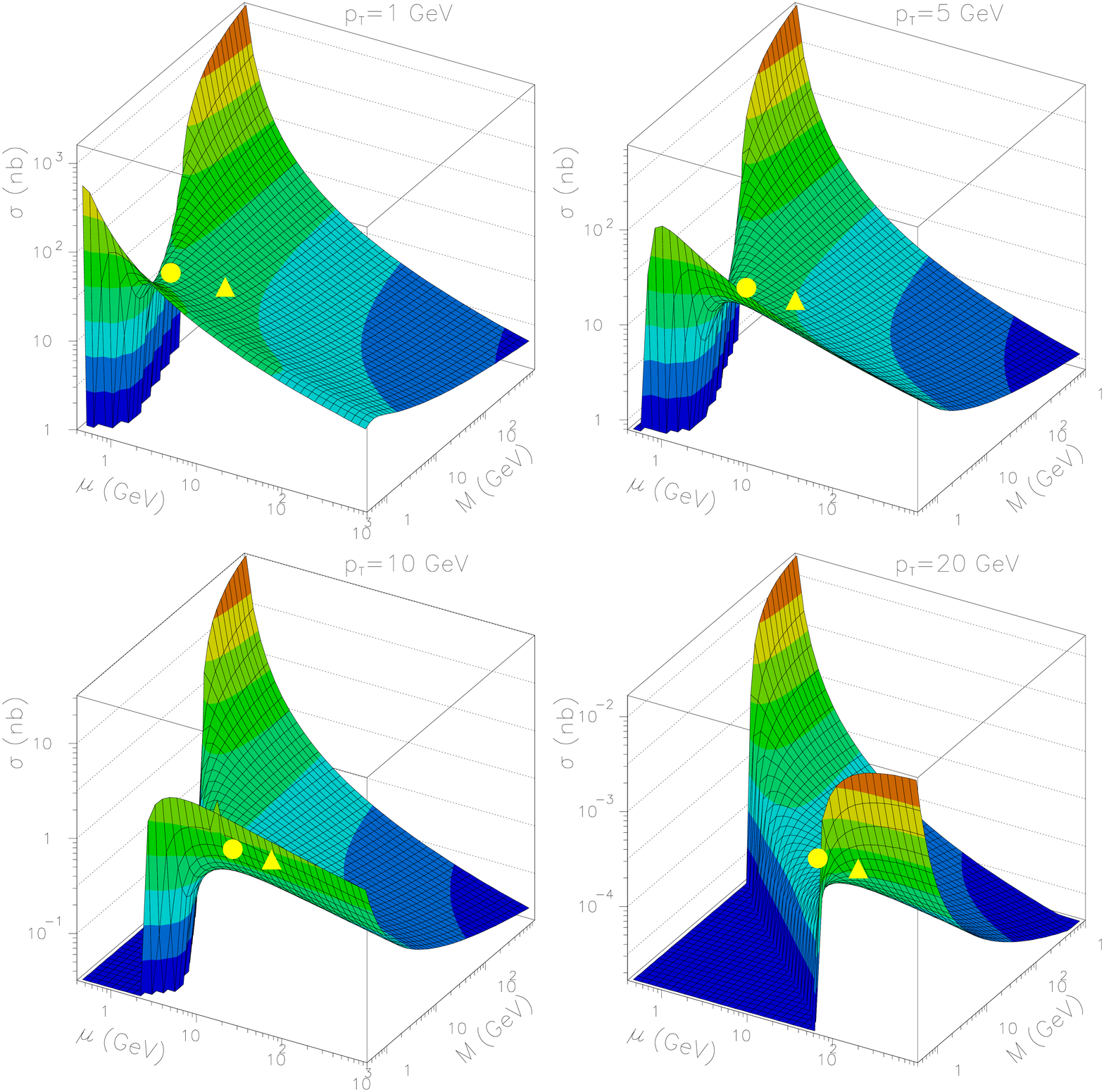}
\caption{(Left) Positions of saddle points, b-quark cross
section and ratio \eqref{ratio} for CTEQ6.1 (green) and GRV94 (black)
PDF at $\sqrt{S}=64$~GeV. Red line shows conventional choice
$\mu=M=\sqrt{p_\perp^2+m^2_b}$. (Right) Surfaces plots for some four
typical values of $p_\perp$, calculated with CTEQ6.1 PDF. Yellow
circles mark saddle points and yellow triangles conventional choice.}
\label{b_64}
\end{figure}

We start with $\overline{\mathrm{p}}$p collisions at $\sqrt{S}=64$ GeV.
In Fig. \ref{b_64} the quantity (\ref{xs}), integrated over the rapidity $y$,
is displayed as a function of $\mu$ and $M$ for several values of $p_\perp$.
On each surface the point corresponding to the standard choice
$\mu=M=\sqrt{p_\perp^2+m_b^2}$ is marked by a yellow triangle. Each of the
surfaces in Fig. \ref{b_64} exhibits a saddle point, marked by a yellow circle.
The $p_\perp$ dependence of the coordinates $(\mu_s(p_\perp),M_s(p_\perp))$ of
these saddle points, displayed in the upper left part of Fig. \ref{b_64}, shows
the first indication of the disparity between the renormalization and
factorization scales: saddle points occur at locations clearly away from the
diagonal $\mu=M$ and always for $\mu_s < M_s$. Moreover, for all values of
$p_\perp$ these saddle points lie almost exactly on the intersection (not
shown) between the LO and NLO surfaces, which defines the points preferred by
the Effective Charge Approach. We shall plot the corresponding contour plot,
illustrating this claim, for the phenomenologically most interesting case of
$\overline{t}t$ production at LHC. In the middle left part of Fig. \ref{b_64}
the distribution ${\mathrm{d}}\sigma/{\mathrm{d}}p_\perp$ computed for the
standard choice $\mu=M=\sqrt{p_\perp^2+m_b^2}$ is compared to that corresponding
to the saddle points. To quantify their difference, we plot in the lower left
part of Fig. \ref{b_64} the ratio (setting $m_Q=m_b$)
\begin{equation}
r(p_\perp)\equiv
\frac{\mathrm{d}\sigma(\mu_s,M_s,p_\perp)/\mathrm{d}p_\perp}
{\mathrm{d}\sigma(\mu=M=\sqrt{p_\perp^2+m_Q^2})/\mathrm{d}p_\perp}.
\label{ratio}
\end{equation}
The difference between the results obtained with CTEQ6.1 and GRV94 PDFs is
negligible as far as the position of saddle points is concerned, but CTEQ6.1
gives, for both the PMS and the conventional choice of scales, slightly higher
values of the cross section. In the region $p_\perp \lesssim 10$ GeV saddle-based
NLO results are by a factor of about $1.5$ higher than the conventional ones.

\begin{figure}[t]
\includegraphics[width=4cm]{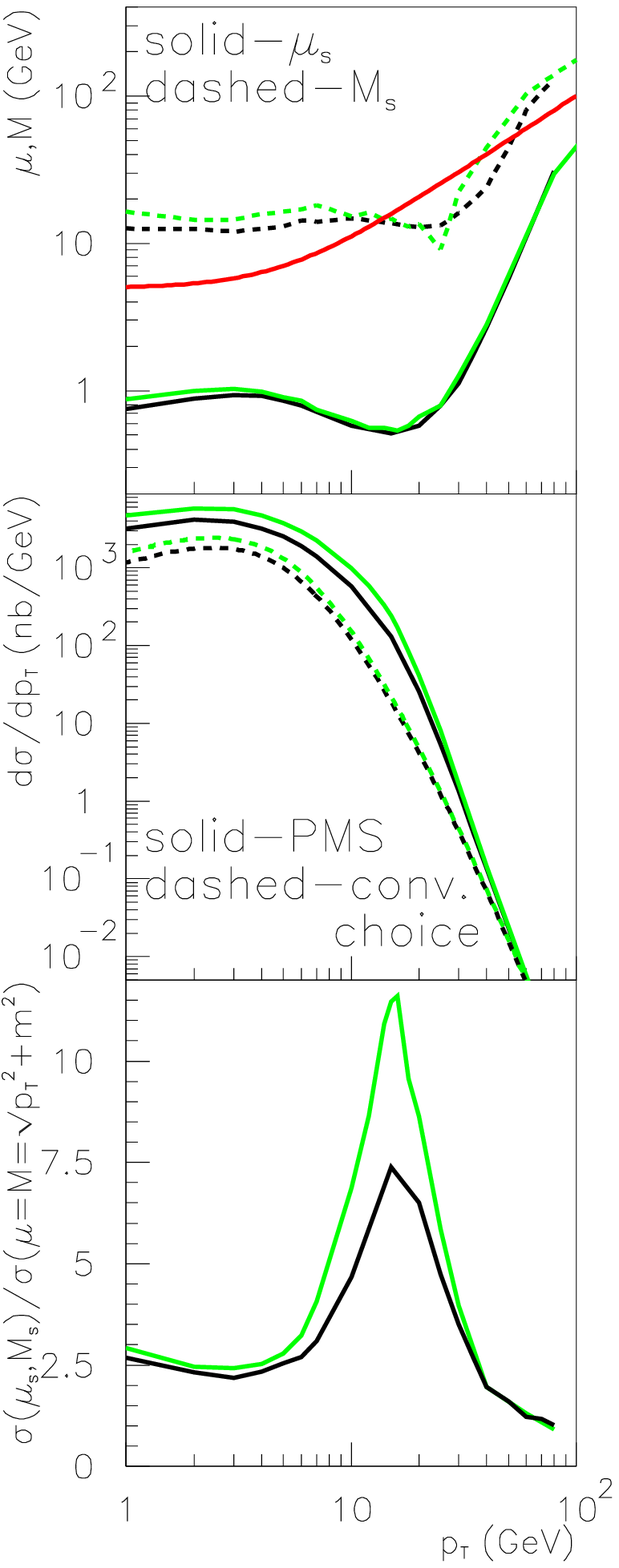}~
\includegraphics[width=11cm]{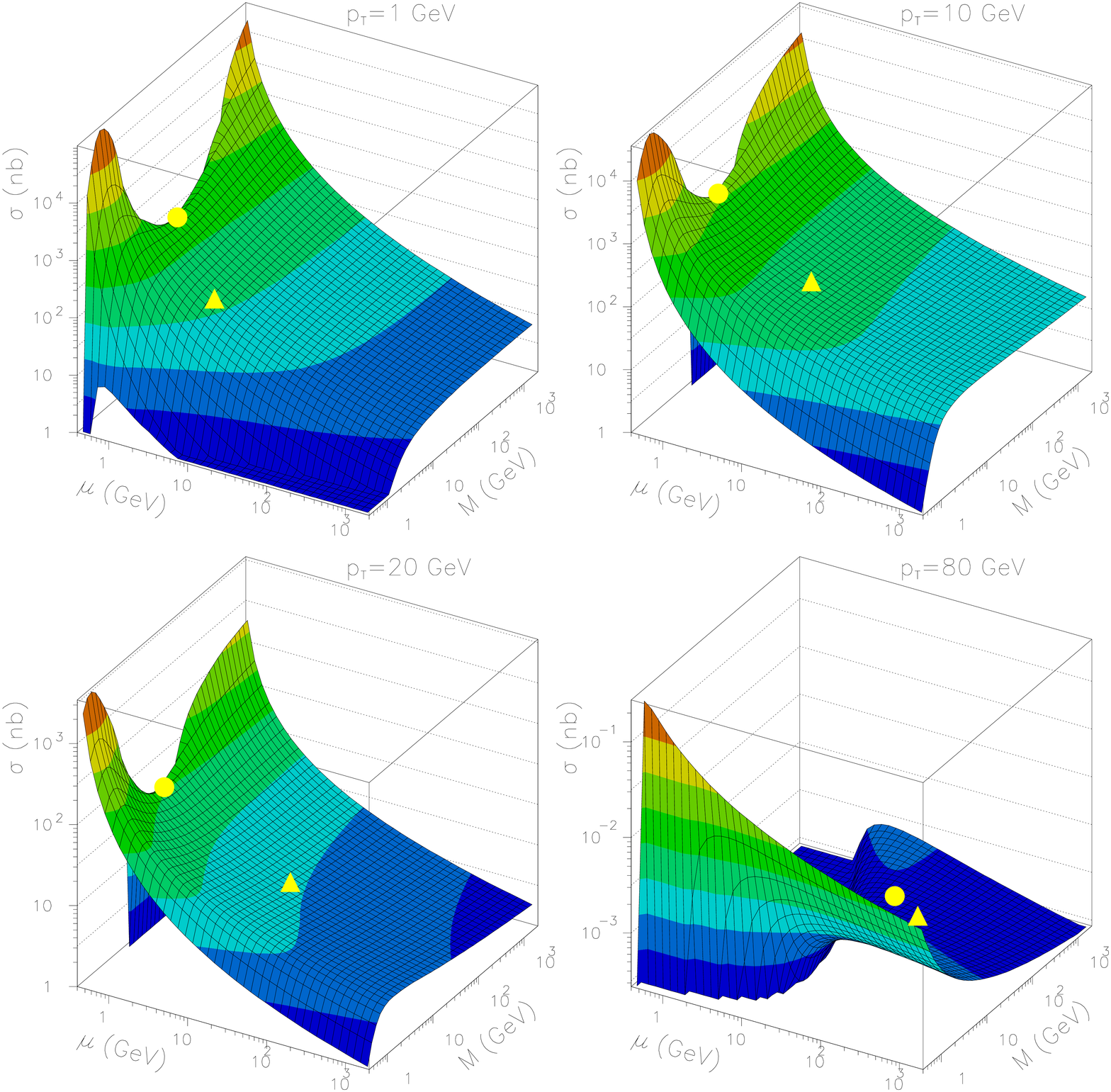}
\caption{The same as in Fig. \ref{b_64} but for $\sqrt{S}=630$ GeV.}\label{b_630}
\end{figure}
As we go to higher energies the relation between the conventional and saddle-based
results changes. At $\sqrt{S}=630$~GeV (see Fig. \ref{b_630}) we find that
\begin{itemize}
\item In the interval $p_\perp \lesssim 100$ GeV the saddle point lies far
away from the diagonal $\mu=M$.
\item For $p_\perp \lesssim 30$ GeV the renormalization scale at the saddle point
$\mu_s\simeq 1$ GeV, dipping to very small value around $0.6$ GeV for
$p_\perp\doteq 16$ GeV, where the ratio (\ref{ratio}) peaks at about 10.
\item Stability of the NLO predictions improves for $p_\perp\gtrsim 100$ GeV,
where the saddle point $(\mu_s,M_s)$ moves closer to the conventional choice
$\mu=M=\sqrt{m_b^2+p_\perp^2}$ and the ratio (\ref{ratio}) approaches unity.
\item
The shapes of the conventional and saddle-based $p_\perp$ distributions differ
substantially, particularly in the region $10\lesssim p_\perp\lesssim 40$ GeV.
\end{itemize}
\begin{figure}[t]
\includegraphics[width=4cm]{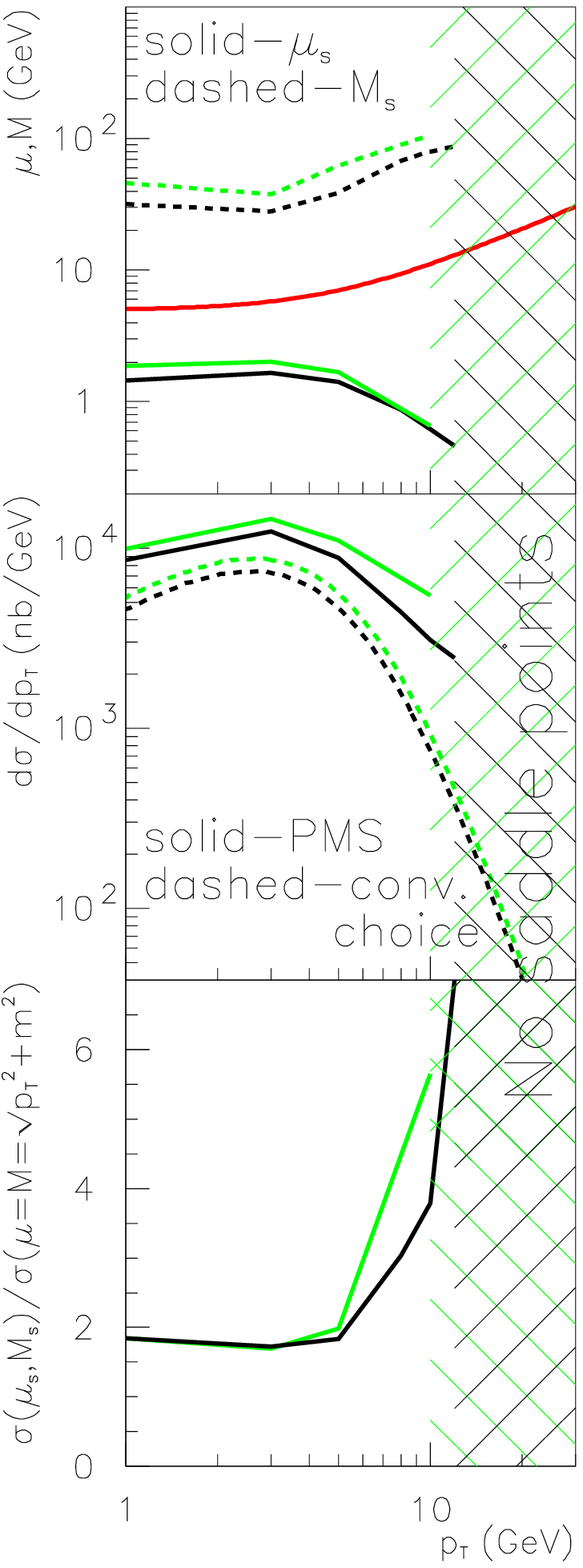}~
\includegraphics[width=11cm]{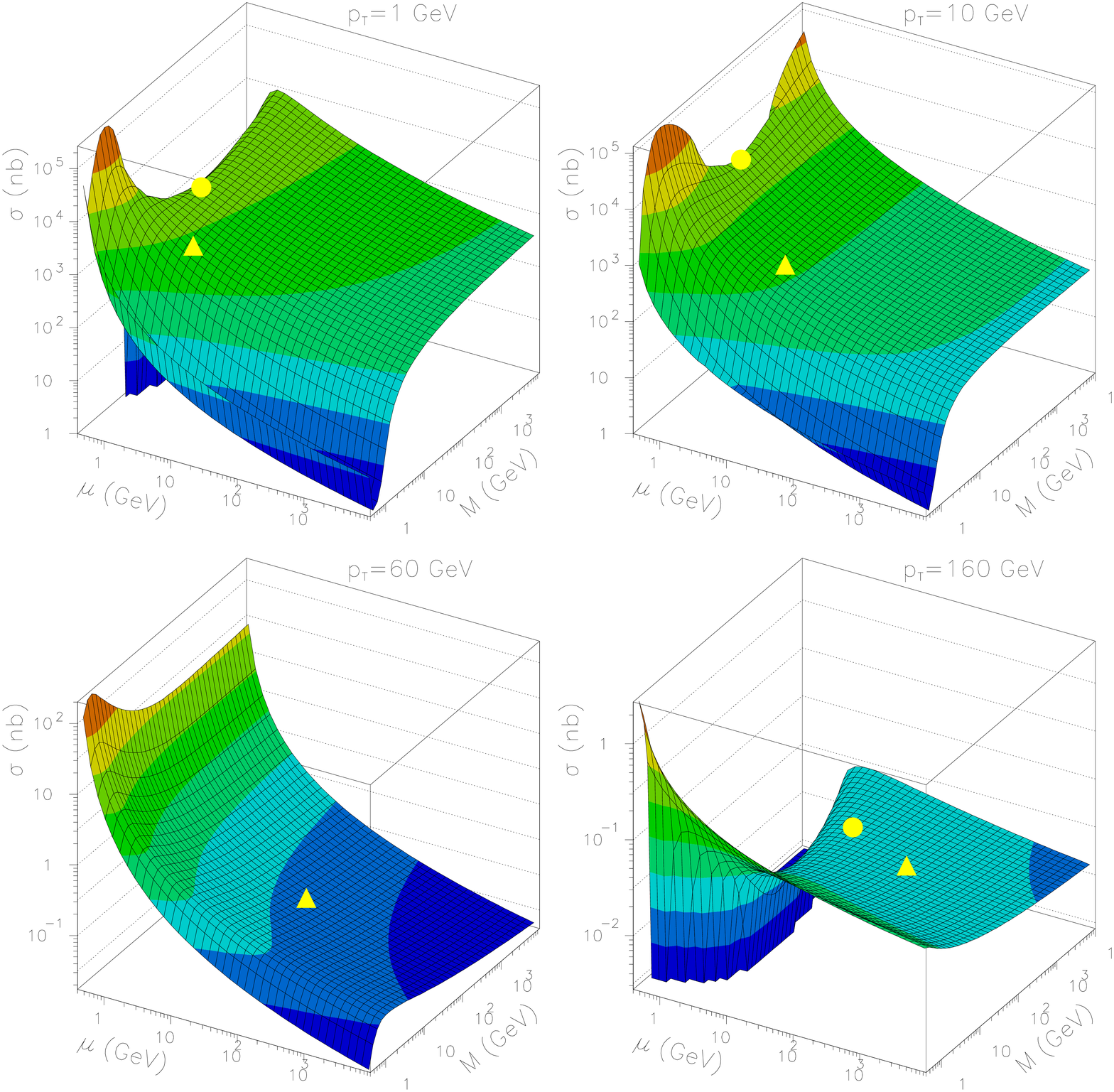}
\caption{The same as in Fig. \ref{b_64} but for $\sqrt{S}=1800$
GeV.} \label{b_1800}
\end{figure}
At this energy and in the experimentally accessible region $p_\perp\lesssim 30$
GeV the typical parton fractions come from the region $x\gtrsim 0.03$ and thus
neither the effects of low $x$ and/or large $p_\perp/m_b$ are likely to change
the fixed order predictions significantly.

For $\overline{\mathrm{p}}$p collisions at $\sqrt{S}=1800$ GeV
(see Fig. \ref{b_1800}) the
saddle point disappears for $p_\perp\gtrsim 10$~GeV and reappears
only for $p_\perp\gtrsim 80$ GeV. The value of the renormalization
scale at the saddle point, $\mu_s$, remains small, dropping below
$1$ GeV as $p_\perp$ approaches $10$ GeV. The saddle point remains
close to the intersection of the LO and NLO surfaces even as the
saddle point approaches the unphysical region, though in this case
the distance between the saddle point and the mentioned intersection
increases. For $p_\perp\lesssim 5$ GeV the ratio (\ref{ratio}) stays
constant at about 2, but then rapidly rises, diverging at
$p_\perp\doteq 10$ GeV. Only for very large and experimentally
inaccessible region $p_\perp\gtrsim 160$~GeV do the conventional
results approach the saddle-based ones. At these large transverse
momenta the resummation of the large logs of the type
$\ln(p_\perp/m_b)$ must certainly be performed to improve the NLO
calculations. However, in most of the $p_\perp$ range accessible at
the TEVATRON, i.e. $p_\perp\lesssim 50$, this effect is likely to be
small. For moderate $p_\perp$ the NLO QCD predictions of inclusive
bottom production are thus highly unstable. For very small
$p_\perp$, roughly $p_\perp\lesssim 10$ GeV, one may argue that
because the relevant parton fractions are typically $0.01$, the
approach \cite{hannes} using the unintegrated PDF within the
$k_T$-factorization may be superior to the present fixed NLO
results. It should, however, be kept in mind that the agreement with
TEVATRON data relies on the use of a particular set of unintegrated
PDF (which are still not well-constrained from data) and the
appropriate choice of the factorization scale. In the case of
$k_T$-factorization there is, however, no mechanism for the
cancelation of the factorization scale dependence between the
unintegrated PDF (which depend on it) and the off-mass-shell hard
scattering cross section as the latter is taken to the LO only and
is thus independent of $M$.

For the proton-proton collisions at the LHC the interval where
the NLO QCD predictions exhibit no stability region and behave as a
monotonously decreasing functions of the renormalization scale $\mu$ extends
between $10\lesssim p_\perp\lesssim 900$ GeV. In the lower part of this range
the partons (mostly gluons) involved in hard collision have momentum fractions
typically of the order of $10^{-3}$, for which the calculations \cite{hannes}
based on unintegrated PDF is likely to be more appropriate. On the other hand,
for large transverse momenta the resummation of logs $\ln p_{\perp}/m_b$ is
definitely necessary to arrive at a reasonably stable result. Compared to these
effects the threshold ones are likely to be negligible.

\subsection{$\overline{t}t$ production}
\label{t}
We have computed
${\mathrm{d}}\sigma/{\mathrm{d}}p_\perp$ for $\overline{\mathrm{p}}$p collisions
at $\sqrt{S}=1800$~GeV and for pp collisions at $\sqrt{S}=14$~TeV, setting in
both cases $m_t=175$~GeV.

\begin{figure}[th]
\includegraphics[width=4cm]{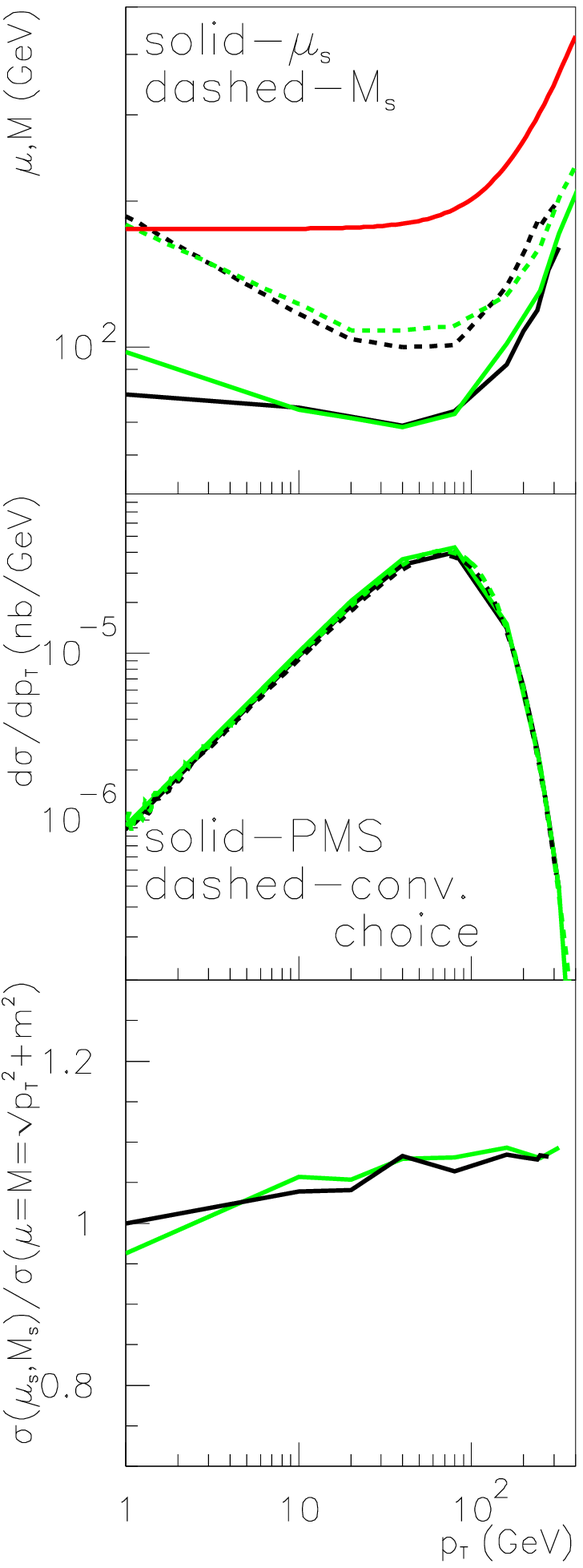}~
\includegraphics[width=11cm]{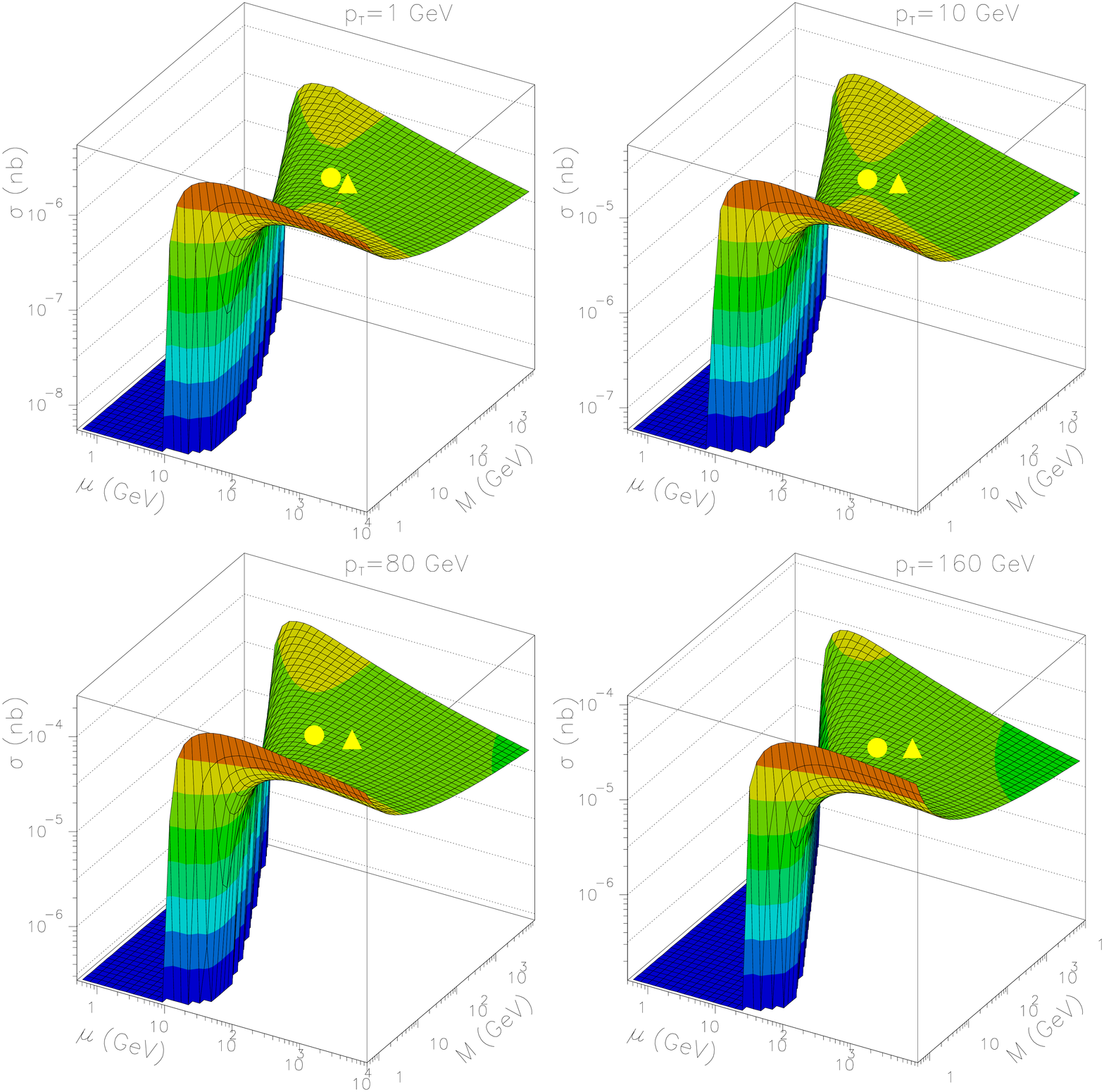}
\caption{The same as in Fig. \ref{b_64} but for $\overline{t}t$ production
at $\sqrt{S}=1.8$ TeV.}
\label{t_1800}
\end{figure}

For the former case the results, presented in Fig. \ref{t_1800}, exhibit saddle
points for $p_\perp\lesssim 200$~GeV. As these saddles are broad, the points
corresponding to the standard choice $\mu=M=m_t$ are not far away and the
saddle-based and standard results are close, typically to within less than $10$ \%.
Moreover, contrary to the case of $\overline{b}b$ production, the saddle points
are located at large renormalization scale $\mu_s\gtrsim 80$ GeV.
At the TEVATRON about $\simeq 85$\% of the cross section comes from light quark
$\overline{q}q$ annihilation with the remaining roughly 15\% from gluon-gluon fusion.
The typical parton fractions $x_q,x_G\gtrsim 0.2$ come from the region where
quark distribution functions are very well (with the precision better than about 3\%)
constrained by existing data, but the gluon distribution function suffers from much
larger uncertainties. Recent error analysis by both CTEQ and MRST groups
\footnote{The numbers quoted below concerning the accuracy of PDF were obtained
with the help of Durham HEPDATA PDF interactive facility \cite{durham} using
CTEQ6 and MRST E-sets.}
show that just above $x_G\simeq 0.2$ this uncertainty starts to increase quite
rapidly, reaching
$30$\% above $x_G\doteq 0.3$ for CTEQ or $x_G\doteq 0.4$ for MRST, respectively.
Despite the fact that gluon-gluon channel contributes just a small part of the
cross section, the overall uncertainty may thus be comparable to or even larger
than that resulting from the scale ambiguity. This fact was noted in
\cite{ttbartev}
\footnote{Where the theoretical uncertainty was defined in a standard manner
by changing the common scale $\mu=M$ by a factor of 2 around $m_\perp$.}
which contains detailed investigation of the dependence of NLO QCD predictions,
including the soft gluon resummation, on the variation of the scales as well as
on choice of parton distribution functions. The soft gluon resummation
reflects the fact that at TEVATRON the $\overline{t}t$ production threshold
is not far below $\sqrt{S}$. Unfortunately, as the calculations of \cite{bonciani},
used in \cite{ttbartev}, are available for the total $t\overline{t}$ cross
section only, we cannot investigate their
implications for the transverse momentum distributions considered in the present
paper. Taking as an estimate of their importance the number given in \cite{bonciani}
for the total cross section $\sigma_{tot}(\overline{p}p\rightarrow \overline{t}t)$,
i.e. the enhancement of the standard NLO results with $\mu=M=m_t$ by about $4$\%,
we expect them to be comparable to the difference between the standard and
saddle-based NLO results. However, as the scale dependence (assuming $\mu=M$) of
the resummed results for $\sigma(\overline{p}p\rightarrow\overline{t}t)$ is reduced
substantially with respect to standard NLO calculations \cite{bonciani}, it would
be quite interesting to see the effects of the soft gluon resummation on the
surfaces in Fig. \ref{t_1800}.
\begin{figure}[th]
\includegraphics[width=4cm]{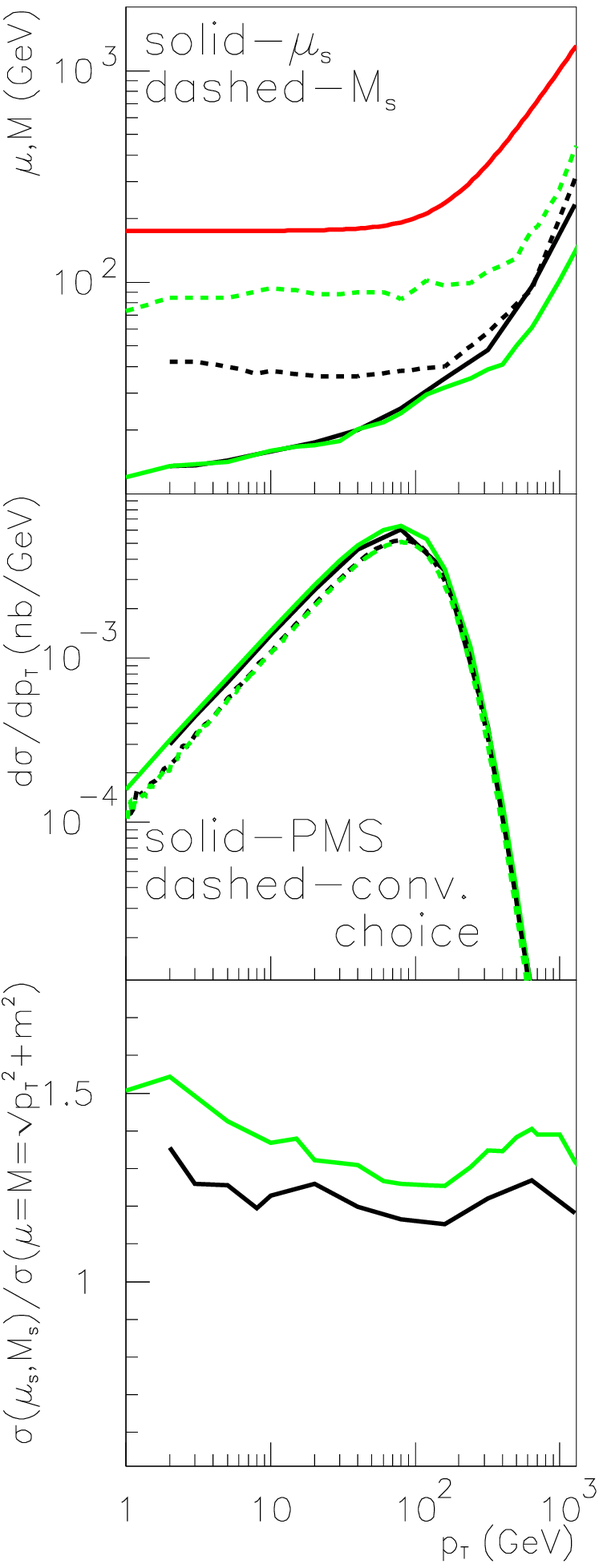}
~\includegraphics[width=11cm]{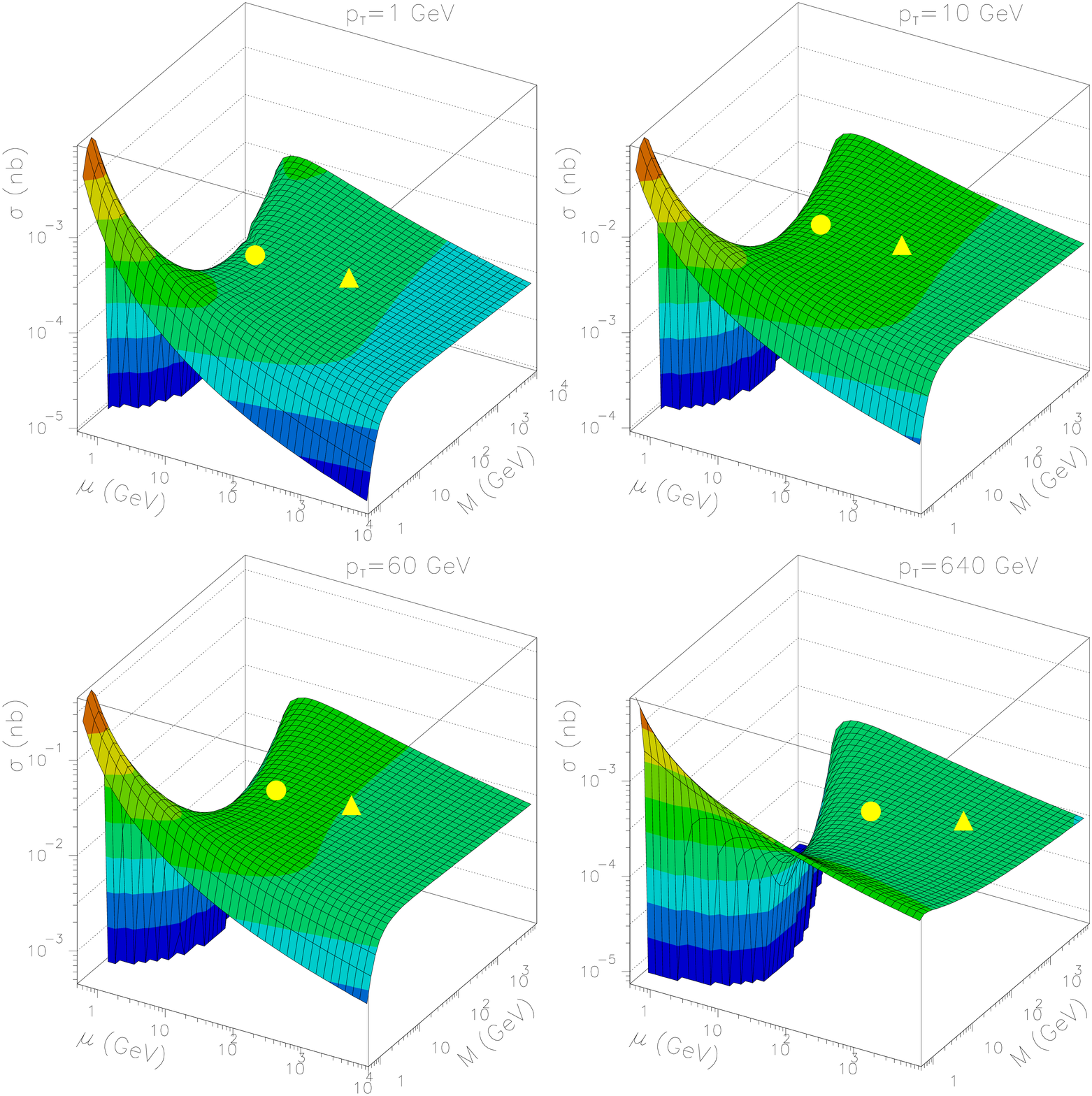}
\caption{The same as in Fig. \ref{b_64} but for $\overline{t}t$ production in
pp collisions at $\sqrt{S}=14$ TeV.}
\label{t_14000}
\end{figure}

Though non-negligible, all the uncertainties of higher order QCD corrections
discussed above are nevertheless substantially smaller than the current experimental
errors even for the total cross section. It is difficult to quote just one number
as the results from various channels and from D0 and CDF experiments differ sizably
(see Figs. 5 and 7 of a recent review \cite{cdftop}), but a conservative estimate
of the experimental uncertainty for the total cross section is $30-50$\%,
depending on the way one adds statistical and systematical errors (the number in
\cite{cdftop} reads $\sigma_{tot}(\overline{p}p\rightarrow\overline{t}t)=
8.2^{+2.4}_{-2.1}(stat)^{+1.8}_{-1.0}(syst)$). The only conclusion one can thus
draw from the analysis of the TEVATRON data on
$\sigma_{tot}(\overline{p}p\rightarrow\overline{t}t)$ is that they are
consistent with QCD calculations, with or without the soft gluon resummation.

At the LHC and even in the initial low luminosity period, $\overline{t}t$ pairs
will be produced abundantly, with millions of such pairs produced per year by
each of the two main experiments. The gluon-gluon channel contributes most of the
cross section, except for very large $p_\perp\gtrsim 700$ GeV, there the light
quark-gluon channel becomes of comparable importance. The $\overline{q}q$ channel
contributes less than $10$\% throughout the experimentally accessible regions.
The precision of the measurements of $\overline{t}t$ production will thus be
limited primarily by the uncertainty of the luminosity, which is expected to be
in the range $5-10$\% \cite{tdr2}.

\begin{figure}\centering
\includegraphics[width=4cm]{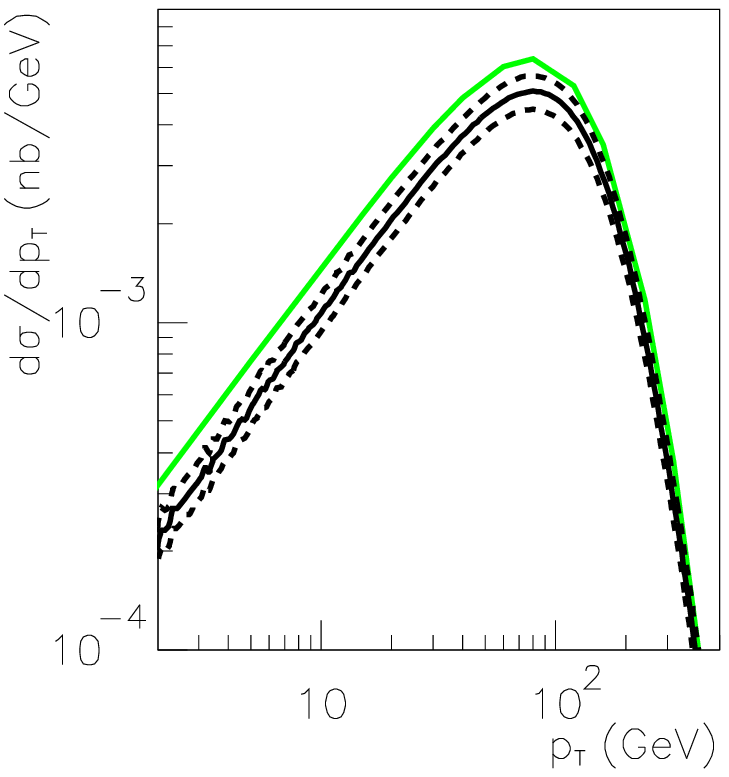}
~\includegraphics[width=8cm]{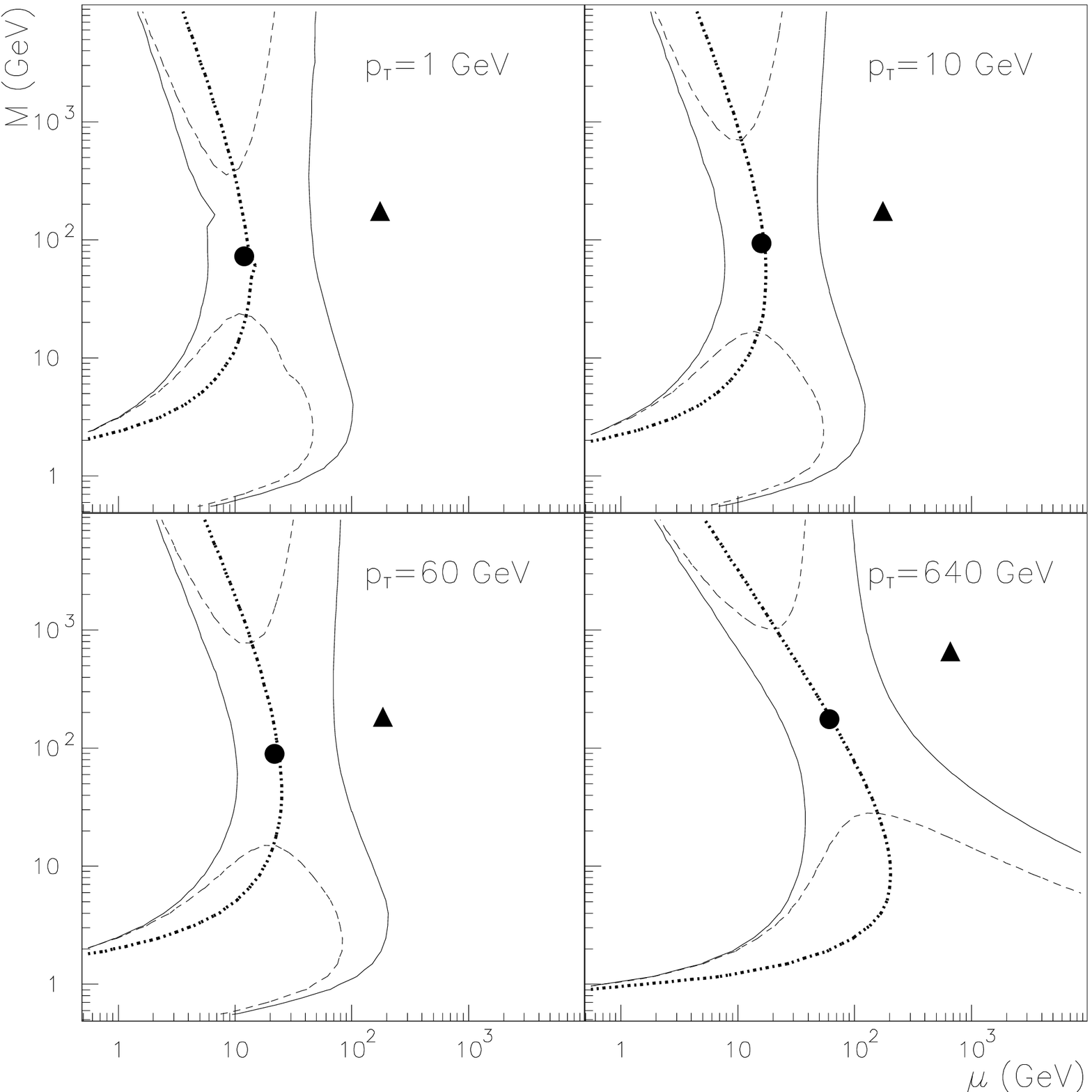}
\caption{Left: the standard way of estimating the uncertainty of NLO calculation
(lower solid black curve) by evaluating the cross section in the band between
$\mu=M=m_T/2$ (upper dotted curve) and $\mu=M=2m_T$ (lower dotted curve), where
$m_T\equiv \sqrt{m_t^2+p_\perp^2}$. The saddle-based results are represented by
the upper solid green curve. Right: contour plot corresponding to the
surface plot in Fig. \ref{t_14000} where the dotted curves show the choice of
scales preferred by the approach of \cite{EC}.}
\label{contours}
\end{figure}

As shown in Fig. \ref{t_14000}, for
$\overline{t}t$ production at the LHC the saddle points exist for all accessible
values of $p_{\perp}$, but they are located further away from the points
corresponding to the standard choice, primarily due to the fact that the
renormalization scale at the saddle points, $\mu_s$, remains below $100$ GeV.
This by itself is not so important, what matters is the fact that the difference
between the saddle-based results and the conventional is quite sizable. For
$p_\perp\lesssim 100$~GeV and the CTEQ6.1 set of PDF the former are higher by a
factor of $1.3-1.5$. This is by about a factor of two more than the upper limit
of the theoretical band based on the standard way of estimating the ``theoretical
uncertainty'' of NLO calculations, shown by the lower (black) solid curve in
the left part of Fig. \ref{contours}.
This enhancement is phenomenologically relevant since at LHC the typical
parton fractions will be by a factor of seven smaller than at the
TEVATRON and thus in the region where both light quark and gluon distributions
functions are known with accuracy better than about 3\% \cite{durham}. The
sensitivity of NLO
QCD predictions to the choice of PDF is thus expected to be substantially smaller
than the difference between the standard and saddle-based NLO calculations, with
or without the soft gluon resummation. The effects of the resummation of large
logs coming from the region of small $x$ and/or large $p_\perp/m_t$ are also
unlikely to be important in the experimentally accessible region.

As demonstrated by the four contour plots in Fig. \ref{contours} the saddle points
are located close to the curves representing the intersections of the LO and NLO
predictions, the feature mentioned already in the case of $\overline{b}b$ production
at $\sqrt{S}=64$ GeV. Note moreover, that for all $p_\perp$ these intersections
follow closely the ridge on both sides of the saddles implying that out of all
the points preferred by the Effective Charges Approach \cite{EC} the one closest
to the saddle gives the lowest value of the cross section.

\section{Summary and conclusions}
We have investigated the renormalization and factorization scale dependence
of inclusive $\overline{b}b$ and $\overline{t}t$ production in pp and
$\overline{\mathrm{p}}$p collisions as a function of $p_\perp$.
For bottom quark at 64~GeV and top quark at TEVATRON and LHC energies, the cross
sections were found to exhibit regions of local stability around the saddle points
for all experimentally relevant values of $p_{\perp}$. In all these cases the
standard choice of scales $\mu=M=\sqrt{p^2_\perp+m^2}$ yields results which are
lower than those at the saddle points.
The difference is smallest for $\overline{t}t$ production at TEVATRON
energie, where it amounts to about $10$\%. For pp collisions at the LHC the
enhancement of the saddle-based predictions over the standard one increases to
to $1.3-1.5$, depending on $p_\perp$. As at LHC top quark production cross
sections will be measured with a precision of about $5-10$\%, and hopefully even
better, theoretical predictions should be of comparable accuracy. In view of this
fact the latter enhancement is phenomenologically relevant. This is
our most important conclusion.

For smaller ratio $m/\sqrt{S}$, characteristic for the bottom production at
TEVATRON and LHC, the saddle points are located far away from the
diagonal $\mu=M$ and the corresponding saddle-based results are substantially
higher that the standard ones. For some regions of $p_\perp$ there is no
region of local stability at all.
For small values of $p_\perp$ the process is dominated by very small partonic
fractions and thus the use of unintegrated PDF may be more appropriate.
For large $p_{\perp}\gtrsim 50$ GeV, on the other hand, the resummation of the
large logs of the type $\ln p_{\perp}/m_b$ should certainly precede any scale
analysis.

\end{document}